\documentstyle[12pt]{article}
\begin{document}

\tolerance 6000
\hbadness 6000

\def\nsection#1{\section{#1}\setcounter{equation}{0}}
\def\nappendix#1{\vskip 1cm\no{\bf Appendix #1}
\def\thesection{#1}
\setcounter{equation}{0}}
\renewcommand{\theequation}{\thesection.\arabic{equation}}

\thispagestyle{empty}
\begin{flushright}
{\bf hep-th/9405193}
\end{flushright}
\vskip 2truecm

\begin{center}

{ \large \bf  LINEAR DIFFERENTIAL EQUATIONS FOR A FRACTIONAL SPIN
FIELD}\footnote{ Revised version of the preprint DFTUZ/92/24,
to be published in J. Math. Phys.}\\
\vskip0.8cm
{ \bf
Jos\'e L. Cort\'es${}^{a,}$\footnote{e-mail: cortes@cc.unizar.es}
 and Mikhail S. Plyushchay${}^{a,b,}$\footnote{e-mail:
mikhail@cc.unizar.es}\\[0.3cm]
{\it ${}^{a}$Departamento de F\'{\i}sica Te\'orica,
Facultad de Ciencias}\\
{\it Universidad de Zaragoza, 50009 Zaragoza, Spain}\\
[0.5ex]{\it ${}^{b}$Institute for High Energy Physics,
Protvino, Russia}}\\[0.5cm]

\vskip2.0cm

                            {\bf Abstract}
\end{center}
The  vector system of linear differential equations for a field with arbitrary
fractional spin is proposed using infinite-dimensional
half-bounded unitary representations of the $\overline{SL(2,R)}$ group.
In the case of $(2j+1)$-dimensional nonunitary representations
of that group, $0<2j\in Z$, they are transformed into equations for
spin-$j$ fields. A local gauge symmetry associated to the vector system
of equations is identified and the simplest gauge invariant
field action, leading to these equations, is constructed.

\newpage

\nsection{Introduction}

A (2+1)-dimensional space-time offers new possibilities which are not
present in any higher dimensional case: due to the Abelian nature of the
spatial rotation group, $SO(2)$, and the topology of many-particle
configuration space, the spin of a relativistic particle can be an
arbitrary real number, and a generalized statistics, intermediate between
Bose and Fermi statistics, is also possible \cite{1}
(see also reviews \cite{2} and
references therein).
The considerable interest to the field models of such particles, called
anyons, is due to their application to different planar physical
phenomena: the fractional quantum Hall effect, high-$T_{c}$
superconductivity and description of the physical processes in the
presence of cosmic strings \cite{3}.

There are several field models realizing anyonic states.  They appear as
topological solitons in the O(3) nonlinear $\sigma$ model or the $CP^{1}$
model with the topological Hopf term \cite{4}, which turns in the low-energy
limit into the $CP^{1}$ model with the Chern-Simons term  \cite{5}.
In the Higgs models with the topological Chern-Simons
term \cite{6,7}, anyons are the electrically charged vortices,
whereas in the models with the topologically massive vector gauge field,
the anyons are the particles directly associated to the matter field \cite{8}.
But, since all these models contain other states, they do not give a minimal
theory of anyons.
In the best known approach, point particles, described by scalar or spinor
fields, are coupled to a U(1) gauge field, the so-called statistical gauge
field, whose dynamics is governed by the Chern-Simons action \cite{9,2}. This
statistical gauge field changes the spin and statistics of particles, but
here it is not clear whether the only effect of the gauge field is to
endow the particle with arbitrary spin or whether residual interactions
are also present.

Therefore, we arrive at the natural question: is it possible to describe
the anyons in a minimal way, without using the statistical Chern-Simons
gauge field?  For the purpose, one can turn over the group theoretical
approach, generalizing ordinary approach to the description of bosonic and
fermionic fields.  Within such an approach, one can work with the
multi-valued representations of the (2+1)-dimensional Lorentz group
$SO(2,1)$ \cite{10}--\cite{12}, or with the definite infinite-dimensional
representations of its universal covering group $\overline{SO(2,1)}$ (or
$\overline{SL(2,R)}$, isomorphic to it) \cite{12}--\cite{17*}.
Though, there is a
close connection between these two possibilities \cite{12}, the problem of
constructing the field actions in the case of using the multi-valued
representations of the Lorentz group is open.  At the same time, different
variants of the free field equations and corresponding  actions were
proposed for fractional spin fields within the framework of the approach
dealing with the infinite-dimensional representations of
$\overline{SL(2,R)}$ \cite{12}, \cite{14}--\cite{16}.
But here mutually connected
problems of second quantization and spin-statistics relation are still
unsolved.  Therefore, strictly speaking, we cannot use the term `anyons'
for such fractional spin fields before establishing the spin-statistics
relation, and it seems important to
continue a search of new equations and corresponding actions for a
fractional spin field in the framework of the group-theoretical approach.

In the present paper we propose new equations for
fractional spin fields, which, in our opinion, have definite advantages with
respect to those from refs. \cite{12},\cite{14}--\cite{16}. They are linear
differential equations,
and the corresponding fields here, unlike those from equations
proposed in refs. \cite{14,16}, carry irreducible representations of
$\overline{SL(2,R)}$. In this sense the proposed equations are similar
to equations for `semions' (i.e. fields with spin $\pm(1/4+n)$,
$n=0,\pm1,...$),
which have been proposed in \cite{13,17},
and generalized to the case of arbitrary spin fields
with the help of the deformed Heisenberg algebra in a recent paper \cite{17*}.
The equations which we shall construct,
have the following property of `universality':
if we choose in them $(2j+1)$-dimensional nonunitary representation of
$\overline{SL(2,R)}$, we will get the equations for a massive field with
integer or half-integer spin $j$. In particular, at $j=1/2$ and $j=1$ these
equations are reduced to the Dirac equation and to the equation for a
topologically massive vector gauge field, respectively. On the other hand, the
choice of infinite-dimensional unitary representation of the discrete type
series, restricted from below or from above, which is the only additional
possibility allowing nontrivial solutions, gives the equations for
a field with fractional (arbitrary) spin. Therefore, the proposed equations
give some link between the ordinary description of bosonic integer and
fermionic half-integer spin fields, and the fields with arbitrary spin.
Moreover, as we shall see, they represent by themselves the first example
of equations which fix the choice of infinite-dimensional
unitary representations of $\overline{SL(2,R)}$ group
for the description of fractional spin fields.
Also, we shall see that the vector system of linear equations satisfy some
identity which,
being a consequence of the choice of irreducible representation
of $\overline{SL(2,R)}$, can be used as a dynamical principle
for the construction of the corresponding gauge invariant field action.

The paper is organized as follows.
In sect. 2 we investigate the equation,  which, in general case,
establishes a mass-spin relation for $(2j+1)$-- or infinite--component fields,
depending on the choice of the corresponding representation of
$\overline{SL(2,R)}$. Except for the case $j=1/2$ and $j=1$,
it does not describe irreducible representations of (2+1)-dimensional
quantum mechanical Poincar${\rm \acute{e}}$ group $\overline{ISO(2,1)}$.
Then, in sect. 3, proceeding from this equation,
we find in a simple way the system of equations
which desribe a relativistic field with arbitrary (fixed) spin and fixed mass,
i.e., an irreducible representation of $\overline{ISO(2,1)}$. A first attempt
in the direction of identifying the field action is pointed out.
Sect. 4 is devoted to the discussion of the results and to
the concluding remarks. Here, in particular, we demostrate that the
proposed equations unambiguosly fix the choice of the representations
of the discrete series
$D^{\pm}_{\alpha}$ of $\overline{SL(2,R)}$
for the description of fractional spin fields, and that
they are the only possible linear vector equations for such fields.

\nsection{Mass-spin equation}
Let us consider the (2+1)-dimensional field equation \cite{14,15}
\begin{equation}
(PJ-\varepsilon \alpha m)\Psi=0,
\label{maj}
\end{equation}
where $\varepsilon=+1$ or $-1$, and $J^{\mu}$ are the generators of the
$\overline{SL(2,R)}$ group, which satisfy the algebra:
\begin{equation}
[J^{\mu},J^{\nu}]=-i\epsilon^{\mu\nu\lambda}J_{\lambda}.
\label{alg}
\end{equation}
Here, a real parameter $\alpha\neq 0$ defines the value of the
$\overline{SL(2,R)}$ Casimir operator:
\begin{equation}
J^{2}=-\alpha(\alpha-1),
\label{cas}
\end{equation}
i.e. in the case $\alpha=-j$, $j>0$ being integer or half-integer,
we suppose the choice of a $(2j+1)$-dimensional irreducible nonunitary
representation $\tilde{D}_{j}$ of $\overline{SL(2,R)}$, whereas in the case
$\alpha>0$ we mean the choice of an irreducible unitary infinite-dimensional
representation of the discrete type series $D^{\pm}_{\alpha}$ of that group
\cite{18}.
These two types of representations are characterized by the following
property: they have a lowest or a highest state annihilated by
the corresponding operator $J_{+}$ or $J_{-}$ (see below) in the case
of representations $D^{\pm}_{\alpha}$, or both such states in the case of
finite-dimensional representations $\tilde{D}_{j}$.

Let us turn over the case of finite-dimensional representations and
first consider the simplest nontrivial
case of the spinor representation
\begin{equation}
J^{\mu}=-\frac{1}{2}\gamma^{\mu},
\label{spinor}
\end{equation}
\begin{equation}
\gamma^{\mu}\gamma^{\nu}=-g^{\mu\nu}+i\epsilon^{\mu\nu\lambda}\gamma_{\lambda},
\ \gamma^{0}=\sigma^{3},\ \gamma^{i}=i\sigma^{i},\ i=1,2,
\label{gamma}
\end{equation}
where $\sigma^{a},$ $a=1,2,3$, are the Pauli matrices. It is this simplest
case that will help us to find the equations we are looking for. Generators
(\ref{spinor}) correspond to
the 2-dimensional irreducible nonunitary representation $\tilde{D}_{1/2}$
with $-\alpha=j=1/2$, and reduce equation
(\ref{maj}) to the (2+1)-dimensional Dirac equation:
\begin{equation}
(P\gamma-\varepsilon m)\Psi=0.
\label{dir}
\end{equation}
The angular momentum operator
\begin{equation}
M_{\mu\nu}=x_{\mu}P_{\nu}-x_{\nu}P_{\mu}+\epsilon_{\mu\nu\lambda}J^{\lambda}
\label{ang}
\end{equation}
is not hermitian  and therefore it is necessary to use the indefinite
`internal' Dirac scalar product
\[
(\Psi_{1},\Psi_{2})=\overline{\Psi}{}^{a}_{1}\Psi_{2}^{a},\quad
\overline{\Psi}=\Psi^{\dagger}\gamma^{0},
\]
to restore hermiticity.

{}From the Klein-Gordon equation
\begin{equation}
(P^{2}+m^{2})\Psi=0,
\label{kle}
\end{equation}
following from (\ref{dir}), we conclude that
in the case  $-\alpha=j=1/2$ initial equation (\ref{maj})
 describes a particle
with mass $M=m$ and spin $s=-\varepsilon/2$, where $s$ is the eigenvalue
of the relativistic spin operator
\begin{equation}
S=-\frac{1}{2\sqrt{-P^{2}}}\epsilon_{\mu\nu\lambda}P^{\mu}M^{\nu\lambda}.
\label{4}
\end{equation}

In the case of the vector representation ($\alpha=-1$),
\begin{equation}
(J_{\mu})^{\alpha}{}_{\beta}=-i\epsilon^{\alpha}{}_{\mu\beta},
\label{j1}
\end{equation}
we have $J^{2}=-2$, and eq.
(\ref{maj}) becomes the equation for the topologically massive
vector field \cite{19}:
\begin{equation}
(-i\epsilon^{\alpha\mu}{}_{\beta}P_{\mu}+\varepsilon mg^{\alpha}{}_{\beta})
\Psi^{\beta}=0.
\label{7}
\end{equation}
{}From (\ref{7}) it follows that $P_{\mu}\Psi^{\mu}=0$,
and that the field $\Psi^{\mu}$ satisfies the Klein-Gordon
equation (\ref{kle}).
Then, using definition (\ref{4}),
we conclude that the field $\Psi^{\mu}$ has spin $s=-\varepsilon$.
$J^{\mu}$ and the angular momentum operator (\ref{ang})
are hermitian with respect to
the obvious indefinite `internal' scalar product
\[
(\Psi_{1},\Psi_{2})=\Psi_{1}^{*\alpha}g_{\alpha\beta}\Psi_{2}^{\beta}.
\]
Putting $\Psi_{\alpha}=\frac{1}{2}
\epsilon_{\alpha\beta\gamma}F^{\beta\gamma},$
$F^{\alpha\beta}=\partial^{\alpha} A^{\beta}-\partial^{\beta}A^{\alpha}$,
we can rewrite eq. (\ref{7}) in the form of equations of motion for the field
strength tensor \cite{19}:
\begin{equation}
(g_{\mu\lambda}\partial_{\nu}+\frac{1}{2}m\epsilon_{\mu\nu\lambda})
F^{\nu\lambda}=0.
\label{8}
\end{equation}

An arbitrary $(2j+1)$-dimensional nonunitary representation
$\tilde{D}_{j}$ of
$\overline{SL(2,R)}$ can be obtained from the corresponding
$(2j+1)$-dimensional representation $D_{j}$, $j=1/2,1,3/2,...$,
of the $SU(2)$ group.
Indeed, let the hermitian operators ${\cal J}^{a}$,
$a=1,2,3$, be the generators of $SU(2)$ group in the representation
$D_{j}$, i.e.
\begin{equation}
[{\cal J}^{a},{\cal J}^{b}]=i\epsilon^{abc}{\cal J}^{c},
\label{su}
\end{equation}
and
\begin{equation}
{\cal J}^{a}{\cal J}^{a}=j(j+1).
\label{casu}
\end{equation}
Then the substitution
\begin{equation}
J_{0}={\cal J}^{3},\quad
J^{i}=-i{\cal J}^{i},\quad i=1,2,
\label{sub}
\end{equation}
gives us the operators $J^{\mu}$ satisfying commutation relations
(\ref{alg}), and condition (\ref{cas}) with $\alpha=-j$.
Then to have an angular momentum operator (\ref{ang}) as a hermitian
one,  it is
necessary to use the corresponding indefinite scalar product,
which we do not write here for the general case.
Note only that representation (\ref{spinor})
is exactly representation (\ref{sub}) for $j=1/2$
if we put ${\cal J}^{a}=\sigma^{a}/2$,
whereas representation (\ref{j1}) is connected with the
corresponding representation (\ref{sub})
with $({\cal J}^{b})^{ac}=i\epsilon^{abc}$  at $j=1$
via the unitary transformation:
\[
U\tilde{J}^{\mu}U^{-1}=J^{\mu},
\]
where we denoted the operators (\ref{sub}) as
$\tilde{J}^{\mu}$, and $U$ is the
unitary diagonal $3\times 3$-matrix with nonzero elements:
$U^{0}{}_{0}=-i$, $U^{1}{}_{1}=U^{2}{}_{2}=1$.

In the case of $j>1$, the corresponding $(2j+1)$-component field $\Psi$
satisfying the equation (\ref{maj}) with $\alpha=-j$, does not describe an
irreducible representation of the (2+1)-dimensional quantum mechanical
Poincar${\rm \acute{e}}$ group $\overline{ISO(2,1)}$.
Indeed, first of all we note that the equation has no nontrivial solutions
in the cases $p^{2}>0$ and $p^{2}=0$ since according to (\ref{sub}),
operators $J^{i}$ and $J^{0}\pm J^{i}$ have no real nonzero eigenvalues.
Therefore, the nontrivial solutions may exist only for $p^{2}<0$.
Then, passing over to the rest frame ${\bf p}={\bf 0}$ via the corresponding
Lorentz transformation, and using the representation
where the operator $J_{0}$ is diagonal, we find the solutions of the equation
(\ref{maj}):
\begin{equation}
\Psi_{r}\propto \delta(p^{0}-\epsilon^{0}M_{\vert r\vert})\delta({\bf p}).
\label{solf}
\end{equation}
Here
$r=-j,-j+1,...,j-1,j$, except for the value $r=0$ for integer
$j$, $\epsilon^{0}=\varepsilon \cdot sign\ r$,
\begin{equation}
M_{\vert r\vert}=m\frac{j}{\vert r\vert}
\label{euc}
\end{equation}
is the mass of the corresponding state, whereas, according to
(\ref{4}), its spin is $s=-\varepsilon\vert r\vert$.
Therefore, we conclude that eq. (\ref{maj})
describes two states with fixed mass $M=m$
and spin $s=-\varepsilon j$ only in the cases when $-\alpha=j=1/2$ and $1$.
These two states differ in their energy signs.
In all other cases eq. (\ref{maj})
describes a set of $2N$ states, where $N=j$ and $N=(2j+1)/2$ for
the cases of integer and half-integer $j$'s, respectively.

Now let us turn to the case of irreducible unitary infinite-dimensional
representations of the discrete series $D_{\alpha}^{+}$ or $D^{-}_{\alpha}$ of
$\overline{SL(2,R)}$. These representations are characterized by the value
of the Casimir operator (\ref{cas}) with $\alpha>0$,
and by the eigenvalues of the operator $J_{0}$:
$j_{0}^{n}=\alpha+n$ and $j^{n}_{0}=-(\alpha+n)$ in these two series,
respectively, where $n=0,1,2,...$ \cite{18}.
In the representation $D^{+}_{\alpha}$ with diagonal operator $J^{0}$,
the matrix elements of $J^{\mu}$ are \cite{15}:
\begin{equation}
J^{0}_{kn}=-(\alpha+n)\delta_{k,n},
\label{9a}
\end{equation}
\begin{equation}
J^{+}_{kn}=-\sqrt{(2\alpha+n-1)n}\cdot\delta_{k+1,n},\ \
J^{-}_{kn}=-\sqrt{(2\alpha+n)(n+1)}\cdot\delta_{k-1,n},
\label{9b}
\end{equation}
where $J^{\pm}=J^{1}\mp iJ^{2}$ and $k,n=0,1,2,....$
The representation $D^{-}_{\alpha}$
can be obtained from (\ref{9a}), (\ref{9b}) through the substitution
\cite{12}:
\begin{equation}
J_{0}\rightarrow -J_{0},\quad
J_{1}\rightarrow -J_{1},\quad
J_{2}\rightarrow J_{2}.
\label{5}
\end{equation}
Here generators $J^{\mu}$ are hermitian with respect to the positive
definite scalar product $(\Psi_{1},\Psi_{2})=\Psi^{*n}_{1}\Psi^{n}_{2}$.

In the cases of the infinite-dimensional representations $D^{\pm}_{\alpha}$,
eq. (\ref{maj}) is a $(2+1)$-dimensional analog of the Majorana equation
\cite{20}, which appears as the equation for the physical subspace in
the model of the relativistic particle with torsion \cite{15}.
Moreover, the mass spectrum (\ref{euc}) appears in that model too:
it is the spectrum of the model in the euclidean space-time.
Note also that the action for the model of the relativistic particle with
torsion, in turn, appeared as the effective
action for a charged  particle interacting with a $U(1)$
statistical gauge field \cite{21}.

Passing over to the rest frame ${\bf p}={\bf 0}$ in the case when $p^{2}<0$,
we find the solutions of this equation:
\begin{equation}
\Psi_{n}\propto \delta(p^{0}-\varepsilon \varepsilon'M_{n})\delta({\bf p}),
\label{solm}
\end{equation}
where $\varepsilon'=+1$ and $-1$ for representations $D^{+}_{\alpha}$ and
$D^{-}_{\alpha}$, respectively.  Their masses and spins are
\begin{equation}
M_{n}=m\frac{\alpha}{\alpha+n},\quad
s_{n}=\varepsilon(\alpha+n).
\label{mn}
\end{equation}
If we take the direct sum of representations,
$D^{+}_{\alpha}\oplus D^{-}_{\alpha}$,
we will have the states with both energy signs in the massive sector
\cite{12}. Majorana equation (\ref{maj}) as well as its
$(~3~+~1~)$-dimensional analog \cite{20}, besides massive solutions
also has massless and tachyonic solutions (see ref. \cite{15}).

To single out the state with highest mass, $M_{0}=m$, and lowest spin,
$s_{0}=\varepsilon\alpha$, and to get rid of massless and tachyonic
solutions, one can supplement  equation (\ref{maj}), linear in $P^{\mu}$,
with the Klein-Gordon equation (\ref{kle}) \cite{12,15,16}. Obviously,
in the case of finite-dimensional representations $\tilde{D}_{j}$
and for the corresponding choice $\alpha=-j$, these
two equations single out two states with mass $M=m$ and spin
$s=-\varepsilon j$, differing in their energy sign.

\nsection{Linear differential equations for fractional spin}

Since eqs. (\ref{maj}) and (\ref{kle}) are completely independent in the case
of representations $D^{\pm}_{\alpha}$ (as well as in the case of
representations $\tilde{D}_{j}$, $j\neq 1/2,1$), they are not very suitable
equations as a basis for constructing the action and quantum theory of the
fractional spin field.
In this section we shall construct
the set of linear differential equations
for the field with arbitrary fractional spin in such a way that
both equations (\ref{maj}) and (\ref{kle}) will appear as a consequence
of them.

To find such equations, let us multiply eq. (\ref{dir}) by an invertible
operator $\frac{1}{2}\gamma^{\mu}$. Then we obtain the vector system of
three equations:
\begin{equation}
L_{\mu}\Psi=0,
\label{3eq}
\end{equation}
with
\begin{equation}
L_{\mu}\equiv (\alpha P_{\mu}-i\epsilon_{\mu\nu\lambda}P^{\nu}J^{\lambda}+
\varepsilon mJ_{\mu}),
\label{leq}
\end{equation}
where $J_{\mu}=-\frac{1}{2}\gamma_{\mu}$ and $\alpha=-\frac{1}{2}$.
These equations are equivalent to eq. (\ref{dir}). Let us show now that in
the general case, i.e. for the choice of any representation $\tilde{D}_{j}$
or $D^{\pm}_{\alpha},$
these equations are equivalent to eqs. (\ref{maj}) and
(\ref{kle}). Indeed, multiplying eq. (\ref{3eq}) by $J_{\mu}$, $P_{\mu}$
and $i\epsilon_{\mu\nu\lambda}P^{\nu}J^{\lambda}$, we correspondingly get:
\begin{equation}
(\alpha-1)(PJ-\varepsilon \alpha m)\Psi=0,
\label{c1}
\end{equation}
\begin{equation}
\left(\alpha(P^{2}+m^{2})+\varepsilon m(PJ-\varepsilon \alpha m)\right)
\Psi=0,
\label{c2}
\end{equation}
\begin{equation}
\left(\alpha(\alpha-1)(P^{2}+m^{2})+(PJ+\varepsilon(\alpha-1)m)(PJ-
\varepsilon \alpha m)\right)\Psi=0.
\label{c3}
\end{equation}
Whence we immediately arrive at the desired conclusion for
$\alpha>0,$ $\alpha\neq1$, and $\alpha=-j$.
As for the case $\alpha=1$, in which eq. (\ref{c1}) disappears, we note
that eqs. (\ref{c2}) and (\ref{c3}) have no nontrivial solutions in the
massless case $P^{2}=0$, and as a result, these two equations also are
equivalent to equations (\ref{maj}) and (\ref{kle}).
Therefore, the vector set of three equations (\ref{3eq})
describes a relativistic field with spin $s=\epsilon\alpha$ and mass $M=m$
for any corresponding choice of irreducible representations
$\tilde{D}_{j}$ or $D^{\pm}_{\alpha}$.

Moreover, by direct verification one can get convinced that
in the general case any two
equations from eqs. (\ref{3eq}) are equivalent to the complete set of three
equations. For example, in the case of representation $D^{+}_{\alpha}$
it can be easily done with the help of the explicit form of the generators
(\ref{9a}), (\ref{9b}). Therefore, the presence of three equations
(\ref{3eq}) gives us a covariant set of linear differential equations
for the description of an arbitary spin field.
As a consequence, we have the following relation
\begin{equation}
R^{\mu}L_{\mu}\equiv 0
\label{gau}
\end{equation}
with
\begin{equation}
R_{\mu}=\left((\alpha-1)^{2}g_{\mu\nu}
-i(\alpha-1)\epsilon_{\mu\nu\lambda}J^{\lambda}+J_{\nu}J_{\mu}\right)P^{\nu},
\label{R}
\end{equation}
which reflects the dependence
of eqs. (\ref{3eq})  in a covariant way.

For completeness, let us write here the commutation relations of the
operators $L_{\mu}$:
\begin{equation}
[L_{\mu},L_{\nu}]=-im\epsilon_{\mu\nu\lambda}\left(L^{\lambda}+
\frac{P^{\lambda}}{m}(PJ-\varepsilon \alpha m)\right),
\label{LL}
\end{equation}
and note, that in the case $\alpha\neq 1$ they can be rewritten in the
following simple form:
$$
[L_{\mu},L_{\nu}]=-im\epsilon_{\mu\nu\lambda}\left(g^{\lambda\rho}+
\frac{P^{\lambda}J^{\rho}}{m(\alpha-1)}\right)L_{\rho}.
$$

As the simplest action leading to the proposed equations (\ref{3eq}),
we can take
\begin{equation}
A=\int {\cal L}d^{3}x, \quad
{\cal L}=\bar{\chi}^{\mu}L_{\mu}\Psi+ \bar{\Psi}L_{\mu}^{\dagger}\chi^{\mu}+
c\cdot\bar{\Psi} (PJ-\varepsilon \alpha m) \Psi,
\label{act}
\end{equation}
where $c$ is an arbitrary real parameter and
$\chi_{\mu}=\chi_{\mu}^{a}$ and
$\bar{\chi}_{\mu}=\bar{\chi}{}^{a}_{\mu}$
are mutually conjugate fields
with index $a$ taking values in the chosen
representation of $\overline{SL(2,R)}$ group.
The variation of the action (\ref{act}) with respect to
$\bar{\chi}{}^{\mu}$ gives equations (\ref{3eq}), whereas the
$\bar{\Psi}$-variation gives
\begin{equation}
L_{\mu}^{\dagger}\chi^{\mu}+ c\cdot (PJ-\varepsilon \alpha m) \Psi = 0,
\label{l1}
\end{equation}
and, besides, we have corresponding equations for the conjugate fields $
\bar{\Psi}$ and $\bar{\chi}^{\mu}.$ Hence, the basic field satisfy the
equations which we want to have, and from (\ref{l1}) we conclude that
$$
L_{\mu}^{\dagger}\chi^{\mu} = 0
$$
for any choice of $c$ \cite{21*}.
 Now it is necessary to get convinced that the
fields $\chi^{\mu}$ and $\bar{\chi}^{\mu}$are pure auxiliary fields.
In the simplest way this can be done within the Hamiltonian formalism
which we hope to present in a future work.

\nsection{Discussion and conclusions}

We have proposed the system of linear differential equations (\ref{3eq})
for a fractional spin field using infinite-dimensional representations
$D^{\pm}_{\alpha}$. They have the form of a covariant (vector) set of
matrix infinite-dimensional equations, from which only two equations
are independent and the presence of the third one allows to have a covariant
set of equations.
One can show that eq. (\ref{3eq}) is in fact the only possible linear
vector set of equations for a fractional spin field.
In other words, if we take an arbitrary linear combination of the operators
$mJ_{\mu}$, $P_{\mu}$ and $\epsilon_{\mu\nu\lambda}P^{\nu}J^{\lambda}$
as the operator $L_{\mu}$ and then demand that equations
of the form (\ref{3eq})
would be equivalent to eqs. (\ref{maj}) and (\ref{kle}), we shall obtain for
the operators $L_{\mu}$ the form (\ref{leq}).

Moreover, the following more general remarkable property of eqs. (\ref{3eq})
is valid. Let us take a set of linear differential
equations of the form (\ref{3eq}) as the equations for a
(2+1)-dimensional field, assuming that
$L_{\mu}=\alpha P_{\mu}
-i\beta\epsilon_{\mu\nu\lambda}P^{\nu}J^{\lambda}+
\varepsilon mJ_{\mu}$,
and that the generators
$J_{\mu}$ are the most general
translation-invariant Lorentz group generators satisfying
the commutation relations (\ref{alg}) (i.e. not fixing the choice of
a representation of $\overline{SO(2,1)}$ from the very beginning).
In this case the parameters $\alpha$ and $\beta$ are arbitrary dimensionless
constants.
Then, multiplying these linear equations by the operators
$mJ_{\mu}$, $P_{\mu}$ and $-i\epsilon_{\mu\nu\lambda}P^{\nu}J^{\lambda}$,
we find that there are only two possible cases in which eqs. (\ref{3eq})
are consistent. The first case is trivial and corresponds to the choice
of a trivial representation for generators: $J_{\mu}=0$, and, therefore,
to a trivial system with $p_{\mu}=0$.
In the nontrivial case there is an arbitrariness in the normalization of the
operator $L_{\mu}$which can be fixed by putting $\beta=1$. Then
the system of vector equations (\ref{3eq}) will be
equivalent to the equations (\ref{maj}), (\ref{kle}) and
\begin{equation}
(J^{2}+\alpha(\alpha-1))\Psi=0.
\label{ir}
\end{equation}
Eq. (\ref{ir}) is simply the condition of irreducibility,
and one can check that the system of eqs. (\ref{maj}), (\ref{kle}) and
(\ref{ir}) is consistent only in the case of the choice of either
finite-dimensional nonunitary representations $\tilde{D}_{j}$, or the
infinite-dimensional unitary representations $D^{\pm}_{\alpha}$.

Therefore, eq. (\ref{3eq}) is the most general vector set of linear
differential equations for a fractional (arbitrary)
spin field in $2+1$ dimensions, whose consistency
fixes the choice of unitary representations of the universal covering
group of (2+1)-dimensional Lorentz group.
 Let us notice here that refs. \cite{12},\cite{14}--\cite{16} have used
representations $D^{\pm}_{\alpha}$ for the description of fractional
spin fields proceeding, in fact, simply from the first quantized theory
of the relativistic point particle with torsion,
and have not excluded for the purpose the choice of other unitary
infinite-dimensional representations of the principal and supplementary
continuous series of the $\overline{SL(2,R)}$ group
(see refs. \cite{18} and a discussion in ref. \cite{22}).
After fixing the representation we have
property (\ref{gau}) as a simple consequence
of the irreducibility condition $J^{2}=-\alpha(\alpha-1)$,
and, as a result, the number of independent equations for the description
of arbitrary spin fields here
is the same as in the spinor-like system of equations
from the recent paper \cite{17*}.

To conclude, let us list some related problems to be solved.

1. We have constructed the simplest action (\ref{act}) leading
to the proposed equations (\ref{3eq}).
The action is invariant with respect to the local transformations:
$\delta \chi_{\mu}=R_{\mu}^{\dagger} \lambda$, $\delta\Psi=0$,
$\lambda$ being an arbitrary field, due to the identity
(\ref{gau}). Therefore, the following question
arises: is it possible to derive the fractional spin field action
from a gauge symmetry principle based on the local transformations
$\delta \chi_{\mu}=R_{\mu}^{\dagger} \lambda$ ? It is not clear
whether more complicated choices of the field action (including the
possibility of having additional auxiliary fields) will be equivalent
to the action (\ref{act}) or whether some new basic ingredient
should be identified in order to understand the formulation of a
free fractional spin system.
It would be interesting to investigate the possible relationship
between the proposed approach to
the description of a fractional spin field and the approach
based on the use of a $U(1)$ statistical gauge field \cite{9}.
Revealing such possible relationship seems very important because
there are some reasons to expect that anyons can occur only in
gauge theories,
or in theories with a hidden local gauge invariance \cite{7}.

2. One can verify that the prescription of a simple substitution
$P_{\mu}\rightarrow P_{\mu}-eA_{\mu}$ in equations (\ref{3eq}) to describe
the interaction with the simplest $U(1)$ gauge (electromagnetic) field
is consistent only in the case of the spinor representation  (\ref{spinor}).
Therefore, the introduction of the interaction of the fractional spin
field with  gauge fields remains an open problem in the present approach.

3. The next interesting problem consists in the construction of a
corresponding singular classical model
of a relativistic particle, whose quantization would lead to equations
(\ref{3eq}) as the equations for the physical states of the system.
Note, that corresponding classical models leading in an analogous way to
equations (\ref{maj}) and (\ref{kle}), and to `semionic' equations
were constructed in refs. \cite{12,16}, and \cite{17}, respectively.

4. The most interesting and intriguing problem within the approach
considered in this paper is the problem of second quantization of the
fractional spin field. The solution of this problem would answer the
question of spin-statistics relation for such fields.  In connection with
this problem, we would like to make two remarks.  First, we note, that the
infinite-component nature of the fractional spin field within the present
approach can be considered as some indication of a hidden nonlocal
nature of the theory, and, therefore, can be treated in favour of an
existence of a spin-statistics relation \cite{7,23}.  Second, let us point out,
that when performing the second quantization of a fractional spin field
within Hamiltonian approach, an infinite number of Hamiltonian constraints
must appear, which are to single out only one physical component (like
$\Psi_{0}$ from eq. (2.21)) from the infinite component basic field
$\Psi_{n}(x)$. This infinite set of constraints should appropriately be
taken into account.

At last, let us point out here that it seems interesting to investigate
the system of (four) linear vector differential equations for
(3+1)-dimensional field in an analogous way, starting from the
generalization of eqs. (3.1) to the (3+1)-dimensional case.

\nsection{Acknowledgements}

This work has been supported by CICYT (Proyecto AEN 90-0030).
M.P. thanks
P.A.~Marchetti, S.~Randjbar-Daemi, D.P.~Sorokin and D.V.~Volkov for
useful discussions.


\begin{thebibliography}{**}
\bibitem{21*}
The constant $c$ can be eliminated
through a rescaling of the fields. Then, there are two
different cases corresponding to $c=0$ and $c=1$,
which have to be
studied separately.
\end{thebibliography}
\end{document}